\documentclass[twocolumn,epsf,showpacs,prd]{revtex4}
\usepackage{graphicx}
\usepackage{dcolumn}
\usepackage{bm}
\begin{document}

\title{Ultra-High Energy Cosmic Rays in a Structured and Magnetized
Universe}
\author{G\"unter Sigl$^a$, Francesco Miniati$^b$,
Torsten~A.~En\ss lin$^b$}
\affiliation{$^a$ GReCO, Institut d'Astrophysique de Paris, C.N.R.S.,
98 bis boulevard Arago, F-75014 Paris, France\\
$^b$ Max-Planck Institut f\"ur Astrophysik,
Karl-Schwarzschild-Str.~1, 85741 Garching, Germany}

\begin{abstract}
We simulate propagation of cosmic ray nucleons above
$10^{19}\,$eV in scenarios where both the source distribution and
magnetic fields within about 50 Mpc from us are obtained from an
unconstrained large scale structure simulation. We find that
consistency of predicted sky distributions with current data above
$4\times10^{19}\,$eV requires magnetic fields of $\simeq0.1\mu$G in our
immediate environment, and a nearby source density of
$\sim10^{-4}-10^{-3}\,{\rm Mpc}^{-3}$. Radio galaxies could provide
the required sources, but only if both high and low-luminosity radio galaxies
are very efficient cosmic ray accelerators. Moreover,
at $\simeq10^{19}\,$eV an additional isotropic flux component,
presumably of cosmological origin, should dominate over the local flux
component by about a factor three in order to explain the observed isotropy.
This argues against the scenario in which local astrophysical sources
of cosmic rays above
$\simeq10^{19}\,$eV reside in strongly magnetized ($B\simeq0.1
\mu$G) and structured intergalactic medium. Finally we discuss how
future large scale full-sky detectors such as the Pierre Auger project
will allow to put much more stringent constraints on source and
magnetic field distributions.
\end{abstract}


\maketitle

\section{Introduction}

Over the last few years the detection of several giant air showers,
either through ground based detectors~\cite{haverah,agasa} 
or fluorescence telescopes ~\cite{fe,hires}, has 
confirmed the arrival of ultra high energy cosmic-rays (UHECRs)
with energies up to a few hundred EeV 
(1 EeV $\equiv 10^{18}\,$eV).
Their existence poses a serious challenge and is
currently subject of much theoretical research
as well as experimental efforts (for recent reviews see
\cite{reviews,bs-rev,school}).

The problems encountered in trying to explain UHECRs in terms of
``bottom-up'' acceleration mechanisms have been well-documented in a
number of studies (e.g., Refs.~\cite{hillas-araa,ssb,norman}).  In
summary, apart from the specific energy draining interactions in the
source the maximal UHECR energy is limited by the product of the
accelerator size and the strength of the magnetic field. According to
this criterion it turns out that it is very hard to accelerate protons
and heavy nuclei up to the observed energies, even for the most
powerful astrophysical objects such as radio galaxies and active
galactic nuclei.

In addition, nucleons above
$\simeq70\,$EeV suffer heavy energy losses due to 
photo-pion production on the cosmic microwave background (CMB) 
--- the Greisen-Zatsepin-Kuzmin (GZK) effect~\cite{gzk} --- 
which limits the distance to possible sources to less than
$\simeq100\,$Mpc~\cite{stecker}. Heavy nuclei at these energies
are photo-disintegrated in the CMB within a few Mpc~\cite{heavy}.
Unless the sources are strongly clustered in our
local cosmic environment, a drop, often called the ``GZK cut-off'' in
the spectrum above $\simeq70\,$EeV is therefore expected~\cite{bbo},
even if the injection spectra extend to much higher energies.
However, the existence of the latter is not established yet
from the observations~\cite{mbo}. In fact, whereas a cut-off
seems consistent 
with the few events above $10^{20}\,$eV recorded by the fluorescence 
detector HiRes~\cite{hires}, it is not compatible with the 
8 events (also above $10^{20}\,$eV) measured by the AGASA ground 
array~\cite{agasa}. The solution of this problem
may have to await the completion of the Pierre Auger project~\cite{auger}
which will combine the two complementary detection techniques
adopted by the aforementioned experiments.

Adding to the problem, there are no obvious astronomical counterparts
to the detected UHECR events
within $\simeq 100\,$Mpc of the Earth~\cite{elb-som,ssb}. At the same
time, no significant large-scale anisotropy has been observed in
UHECR arrival directions above $\simeq10^{18}\,$eV,
whereas there are strong hints for small-scale clustering:
The AGASA experiment has observed five doublets and one triplet within
$2.5^\circ$ out of a total of 57 events detected above 40 EeV~\cite{agasa}.
When combined with three other ground array experiments, these numbers increase
to at least eight doublets and two triplets within $4^\circ$~\cite{uchihori}.
This clustering has a chance probability of less than
$1\%$ in the case of an isotropic distribution.

Independent of the specific UHECR production mechanism, there are
currently two possible explanations of the experimental findings
described above: The first assumes very weak intergalactic magnetic
fields capable of deflecting UHECRs only up to a few degrees, or
neutral primaries. In this case the apparent isotropy would indicate
that many sources contribute to the observed flux and most of
these sources would be at cosmological distances because the
local source distribution is in general too anisotropic to be consistent
with the observed UHECR isotropy. This would also explain the absence
of nearby counterparts and a subset of especially powerful sources
would explain the small-scale clustering~\cite{tt}. Indeed, it has been
argued that UHECR arrival directions correlate with the positions of
BL Lacertae objects, suggesting these as sources accelerating
protons~\cite{bllac}, although there seems to be disagreement about
this in the literature~\cite{bllac1}. Furthermore, some of these
objects may be too far away to be consistent with the GZK effect,
which would require new physics such as Lorentz symmetry
violations~\cite{jlm}. In contrast, correlations with compact radio
quasars have not been found~\cite{c_quasar}. If correlations with
astrophysical objects are confirmed, this would strongly suggest
small deflection or neutral primary particles. Whatever the sources
are in this scenario, for small deflection one can in principle constrain
the characteristics of the magnetic fields along the line of sight
and the source properties by analysing arrival times, directions,
and energies of observed small-scale multi-plets~\cite{sl}. Also,
in the small deflection scenario
the experimental confirmation of a GZK cutoff is expected.

However, the assumption of weak intergalactic magnetic fields 
seems at odds with several observations~\cite{kronberg}.
Most remarkable are the detections of Faraday rotation measures
which seem to indicate field strengths at the $\mu$G level
within the inner region ($\sim $ central Mpc) of galaxy clusters
~\cite{bo_review}. In addition, the recent mounting evidence for 
diffuse radio-synchrotron emission in numerous galaxy clusters 
\cite{gife00} and in a few cases of filaments 
\cite{kkgv89,bagchietal02}, seems to suggest the
presence of magnetic fields as strong as 0.1-1.0$\mu$G at the
relatively low density outskirts of collapsed cosmological
structures. In fact, extragalactic
magnetic fields (EGMF) as strong as $\simeq 1 \mu G$ in sheets and
filaments of the large scale galaxy distribution, such as in our Local
Supercluster, are compatible with existing upper limits on Faraday
rotation~\cite{bo_review,ryu,blasi}.
It is also possible that fossil cocoons of former radio galaxies,
so called radio ghosts, contribute to the isotropization of UHECR
arrival directions~\cite{mte}.
Thus, relatively strong magnetic fields seem to be ubiquitous in 
intergalactic space, although their theoretical understanding
is still limited~\cite{bt_review}.

Such observational evidence motivates a second, more realistic
scenario, which takes into account the existence of strong 
($B\sim0.1-1\,\mu$G) intergalactic magnetic fields correlated with
the large scale structure. In this case magnetic deflection of
charged primaries would be considerable even at the
highest energies and the observed UHECR flux could be dominated by
relatively few sources within about 100 Mpc. Here, large scale isotropy
could be explained by considerable angular deflection leading to diffusion
up to almost the highest energies and the small scale clustering
could be due to magnetic lensing~\cite{hmrs}. The locations of clusters of
events of different energies would in this case coincide with the
crossing points of the caustics for these energies where fluxes
are enhanced.

In the present paper we take this second point of view and investigate 
in some detail the effects of propagation of UHECRs, assumed to be
dominantly nucleons, in a magnetized
large scale structure matter distribution computed according to 
a numerical cosmological simulation.
 
Early investigations of this scenario have been carried out in
Refs.~\cite{slb,ils,lsb,sse,is}, assuming that sources and magnetic fields
follow a pancake profile of scale height $\simeq3\,$Mpc and scale
length $\simeq20\,$Mpc, the magnetic field having a power law spectrum
at length scales below $\simeq1\,$Mpc. UHECR propagation
was computed through a numerical code that accounts for magnetically
induced deflections and all relevant energy losses~\cite{slb,lsb,ils}. 
The cases of a single 
source~\cite{slb,ils}, as well as continuous~\cite{lsb} and 
discrete source distributions ~\cite{is} have been investigated. 
The above studies led to the result that the multi-pole moments and
autocorrelation functions of the arrival directions best fit
the AGASA data for a number $\sim$10 sources in
the Local Supercluster, assumed to emit continuously, and a maximal
field strength of $\simeq0.3\mu\,$G~\cite{is}.

Ideally, however, it would be desirable to study the 
propagation of UHECRs based on distributions of both potential 
sources and observed magnetic field properties.
However, up to now, only catalogs of candidate sources have been 
available. Magnetic fields, on the other hand, have been approximated
in a number of fashions: as negligible~\cite{sommers},
as uniform~\cite{ynts}, or as organized in spatial cells 
with a given coherence length and a strength
depending as a power law on the local density~\cite{tanco}.

In the present paper we attempt to go beyond some of the 
above limitations by computing for the first time the 
propagation of the UHECRs in a magnetized cosmological 
environment computed through numerical simulations.
We carry out a fully cosmological simulation of large
scale structure formation which, in addition to dark matter 
and baryonic gas, follows the evolution of a passive magnetic 
field. This approach is motivated by the fact that 
$\mu$G magnetic fields are mostly negligible for the purpose 
of the dynamics of the large scale cosmic flows (hence their
passive character). In addition, and basically for the same 
reason, the structure of magnetic fields on scales of interest 
for UHECR propagation ($\sim$ 100 kpc) is mostly determined by 
the hydrodynamic flow. This is confirmed by the fact that in these 
simulations, the magnetic field looses memory of its initial
conditions, soon after the formation of structures begins.
Finally, the statistical properties of 
cosmological structure in the universe are rather homogeneous.
Therefore, the simulated matter structure and magnetic field
distributions should provide a realistic scenario for studying 
the statistical properties of UHECR source distributions and
propagation in a cosmic environment. In the present study we
assume the sources to follow the baryon density. Furthermore,
the observer is supposed to be in regions of the simulated
matter distribution which contain structures of the
same size and baryonic gas temperature as our local neighborhood.
This should provide a suitable environment to simulate the arrival
of UHECRs from extragalactic distances and the effects of local 
magnetic fields of various strengths.

In the future such studies can be further improved by computing
{\it constrained} simulations that reproduce in detail the 
observed matter distribution of the local universe. Such a
simulation has been used for the case of radio ghosts in
Ref.~\cite{mte} where, however, the magnetic fields were
not followed but were rather assumed to scale with the
gas density. Constrained simulations including magnetic
fields are relevant for predicting quantitative features such as
location of clustered events, phases of anisotropies etc. and
will be used in a following study. We point out, however,
that for the reasons given above, effects of the magnetic field and
source distributions in the local universe should 
essentially be captured by the present approach at least up to
``cosmic variance''. The latter represents variations due to
different source and observer locations and will be estimated
in our simulations.

We also restrict ourselves to UHECR nucleons, and we neglect the
Galactic contribution to the deflection of UHECR nucleons since
typical proton deflection angles in galactic magnetic fields of
several $\mu$G are $\lesssim10^\circ$ above
$4\times10^{19}\,$eV~\cite{medina}, and thus in general are small
compared to extra-galactic deflection in the scenarios studied
in the present paper.

The simulation is described in more detail in the next section.  
There we also describe the general features of our method
and define the statistical quantities used for
comparison with the data. In Sect.~3 we present results and we
conclude in Sect.~4.

\section{Motivation and Outline of the Numerical Model}

\subsection{Magnetic Deflection}

Contrary to the case of electrons, for charged hadrons deflection is
more important than synchrotron loss in the EGMF. To get an impression
of typical deflection angles one can characterize the EGMF by its
r.m.s. strength $B$ and a coherence length $l_c$.  If we neglect
energy loss processes for the moment, then the r.m.s. deflection angle
over a distance $r\gtrsim l_c$ in such a field is
$\theta(E,r)\simeq(2rl_c/9)^{1/2}/r_L$~\cite{wm}, where the Larmor
radius of a particle of charge $Ze$ and energy $E$ is $r_L\simeq
E/(ZeB)$. In numbers this reads
\begin{eqnarray}
  \theta(E,r)&\simeq&0.8^\circ\,
  Z\left(\frac{E}{10^{20}\,{\rm eV}}\right)^{-1}
  \left(\frac{r}{10\,{\rm Mpc}}\right)^{1/2}\nonumber\\
  &&\hskip1cm\times\left(\frac{l_c}{1\,{\rm Mpc}}\right)^{1/2}
  \left(\frac{B}{10^{-9}\,{\rm G}}\right)\,,\label{deflec}
\end{eqnarray}
for $r\gtrsim l_c$. This expression makes it immediately obvious
why a magnetized Local Supercluster with fields of fractions
of micro Gauss prevents a direct assignment of sources in
the arrival directions of observed UHECRs; the deflection
expected is many tens of degrees even at the highest energies.
This goes along with a time delay 
\begin{eqnarray}
\tau(E,r)&\simeq&r\theta(E,d)^2/4\label{tau}\\
&\simeq&1.5 \times10^3\,Z^2
\left(\frac{E}{10^{20}\,{\rm eV}} \right)^{-2}
\left(\frac{r}{10\,{\rm Mpc}}     \right)^{2}\nonumber\\
&&\hskip1cm\times\left(\frac{l_c}{\rm Mpc}\right)
\left(\frac{B}{10^{-9}\,{\rm G}}  \right)^2 \, {\rm yr}\,,\nonumber
\end{eqnarray}
which may be millions of years. A source visible in UHECRs today
could therefore be optically invisible since many models involving,
for example, active galaxies as UHECR accelerators, predict
variability on much shorter time scales.

\subsection{Numerical Simulation of the Large Scale Structure}

The formation and evolution of the
large scale structure is computed by means of an
Eulerian, grid based Total-Variation-Diminishing
hydro+N-body code \cite{rokc93}.
We adopt a canonical, flat $\Lambda$CDM cosmological model
with a total mass density $\Omega_m=0.3$ and a
vacuum energy density $\Omega_\Lambda= 1- \Omega_m= 0.7$.
We assume a normalized Hubble constant
$h_{67}\equiv H_0/67$ km s$^{-1}$ Mpc$^{-1}$ = 1
and a baryonic mass density, $\Omega_b=0.04$.
The simulation is started at redshift $z\simeq 60$
with initial density perturbations generated as a Gaussian 
random field and characterized by a power spectrum with a 
spectral index $n_s=1$ and ``cluster-normalization'' 
$\sigma_8=0.9$. 

We adopt a computational box size of $50\,h_{67}^{-1}\,$Mpc. In
this box the dark matter component is described by 256$^3$ particles
whereas the gas component is evolved on a comoving grid of 512$^3$
zones. Thus each numerical cell measures about $100\,h_{67}^{-1}\,$kpc
(comoving) and each dark
matter particle corresponds to $2\times 10^9\,h_{67}^{-1}\,M_\odot$.
Besides the box and dark matter particle sizes the 
cosmological simulation is the same as that presented in Ref.~\cite{miniati}.

\begin{figure}
\includegraphics[width=0.45\textwidth,clip=true]{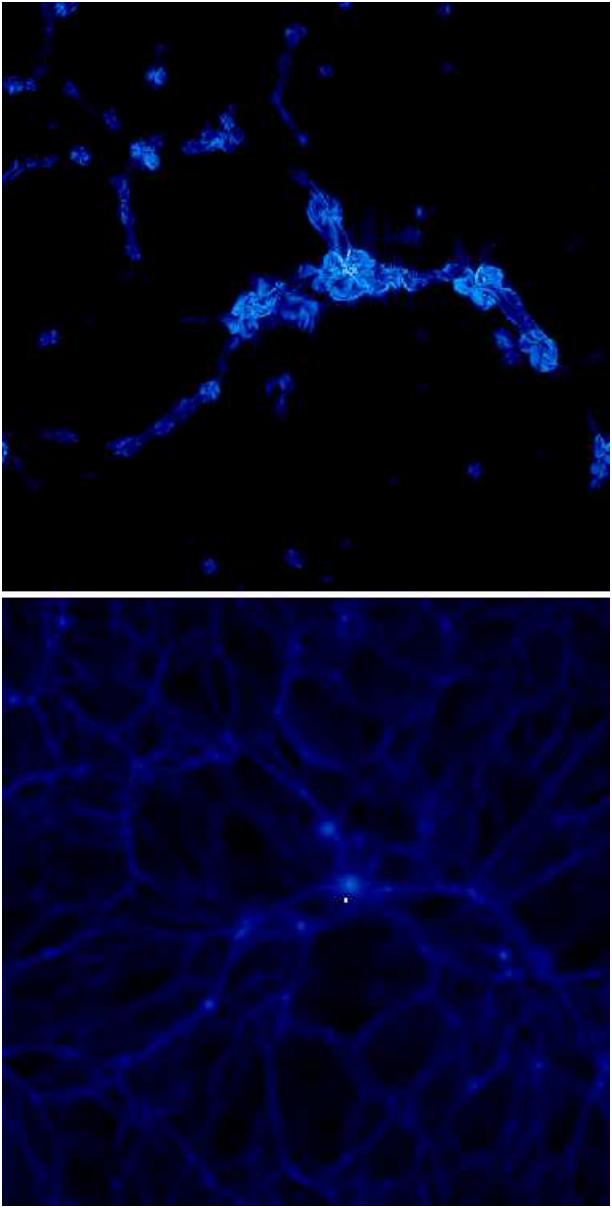}
\caption[...]{Log-scale two-dimensional cut through 
magnetic pressure (top) and baryon density (bottom). The image 
is $50\,h_{67}^{-1}\,$Mpc on each side and $100\,h_{67}^{-1}\,$kpc deep.
The small white dot in the bottom panel indicates the location 
of the observer.
For visualization purposes we adopt a dynamic range of 3 and 
6 orders of magnitude for the magnetic pressure and baryon 
density, respectively.}
\label{fig1}
\end{figure}

The magnetic field is followed as a passive quantity, that is 
magnetic forces are neglected. This is consistent with the 
strength of observed magnetic fields in most diffuse 
extragalactic environments.
Basically we solve the induction equation with the velocity
field provided by the simulated flow \cite{kcor97} and the initial
magnetic field seeds generated by the Biermann battery mechanism.
However, as already pointed out, the initial conditions are 
not important as the topological properties of the 
magnetic field are determined by the subsequent evolution of
the large scale flow. This is responsible for its amplification
through gas compression and shear flows.
Thus, at the end of the simulation, the relative strength of 
the magnetic field in different regions is determined by the
hydrodynamic properties of the flow.
While the simulation outcome regarding the {\it relative} magnetic 
field strength and topology distribution are obviously retained, 
the overall normalization is chosen in order to reproduce the
fields of several micro Gauss observed in the regions of largest
density, namely galaxy cluster cores. Fig.~\ref{fig1} illustrates 
an example of the simulated magnetic pressure (top) and 
baryonic density (bottom) distributions. The figure 
shows two-dimensional cuts corresponding to a depth of 
$100\,h_{67}^{-1}\,$kpc. The color images are in log 
scale and, for visualization purposes, span a 
dynamic range of 3 and 6 orders of magnitude for magnetic pressure
and baryonic density respectively. The magnetic field 
is particularly strong in both postshock regions and inside 
relatively large structures where it has been compressed and 
stretched. Apparently, its distribution is less concentrated 
than the baryonic density, resembling in this respect that of 
the thermal pressure (not shown).

\subsection{Simulated UHECR Experiments}

To simulate the propagation and arrival of UHECRs in the 
computational box we need to choose: (a) the location of the 
observer and (b) the source distribution.
As anticipated in the introduction, the location of the 
observer is identified as a region whose general features
in terms of scale, mass and temperature,  
resemble those of the local universe. That means a small 
group of galaxies characterized by a gas temperature of 
order of a fraction of a keV. There are several such structures
in a $50\,h_{67}^{-1}\,$Mpc box such as the one employed here.
In the neighborhood of the one we selected as the observer
location, we also find a larger group of galaxies with 
temperature of a few keV. 
In order to orient the simulation box 
with respect to the observed sky, the latter object,
located at a distance of $\sim$ 34 Mpc, is arbitrarily 
associated with the Virgo cluster.
This reference frame allows us to define a celestial system of 
coordinates $(\alpha,\delta)$ which describes the arrival direction
of events recorded by our virtual observer. It
will be useful in the next section where the arrival
direction probability distribution is constructed.
The above setting is sufficient for the current purpose of 
investigating the effects on the propagation of UHECRs 
of realistic, topologically structured magnetic fields 
of various strengths.

We then chose randomly a certain total number $N_s$ of sources
in the box, corresponding to an average source density
$8\times10^{-6}h_{67}^3\,N_s\,{\rm Mpc}^{-3}$, with 
probability proportional to the local baryon density. 
In order to avoid introducing too many free parameters, we further
assume that all sources roughly emit the 
same power law spectrum of CRs extending up to $\simeq10^{21}\,$eV,
with roughly equal total power. We also assume that neither
total power nor the power law spectral index
change significantly on the time scale of 
UHECR propagation. This can be up to a few Giga years for 
the magnetic fields considered here.
Injected power and spectral index are then treated as parameters which
can be fit to reproduce the observed spectrum, as will be seen below.

For each such configuration many nucleon trajectories originating from the
sources were computed numerically by solving the equation of motion
for the Lorentz force and checking for pion production every fraction
of a Mpc according to the total interaction rate with the CMB and,
in case of an interaction, by randomly selecting the secondary energies
according to the differential cross section. Pair production by
protons is treated as a continuous energy loss process.

A detection event was registered and its arrival directions and 
energies recorded each time the trajectory of the propagating particle
crossed a sphere of radius 1 Mpc around the observer. For each
configuration this was done until 5000 events where registered.
For more details on this method see Refs.~\cite{slb,lsb,ils}.

\subsection{Data Processing}

For each realization of sources and observer, these events were used
to construct arrival direction probability distributions,
taking into account the solid-angle dependent exposure function
for the respective experiment and folding over the angular resolution.

For the exposure function $\omega(\delta)$ we use the parameterization
of Ref.~\cite{sommers} which depends only on declination $\delta$,
\begin{eqnarray}
\omega(\delta)&\propto&\cos(a_0)\cos(\delta)\sin(\alpha_m)+
        \alpha_m\sin(a_0)\sin(\delta)\,,\nonumber\\
&&\mbox{where}\hskip0.5cm\alpha_m = \left\{ \begin{array}{ll} 
\ 0 & \mbox{if $\xi > 1$}\\
\ \pi & \mbox{if $\xi < -1$}\\
\ \cos^{-1}(\xi) & \mbox{otherwise}
 \end{array} \right.\,,\label{expos}\\
&&\mbox{with}\hskip0.5cm\xi\equiv
\frac{\cos(\theta_m)-\sin(a_0)\sin(\delta)}{\cos(a_0)\cos(\delta)}\,.
\nonumber
\end{eqnarray}
For the AGASA experiment $a_0=-35^\circ$, $\theta_m=60^\circ$, and
the angular resolution $2.4^\circ$ are used. For a full-sky Pierre
Auger type experiment we add the exposures for the Southern Auger
site with $a_0=-35^\circ$ and a putative similar Northern site with
$a_0=39^\circ$, and $\theta_m=60^\circ$ in both cases,
with an assumed angular resolution of $\simeq1^\circ$.

From the distributions obtained in this way typically 1000
mock data sets consisting of $N$ observed events were selected randomly.
For each such mock data set or for the real data
set we then obtained estimators for the spherical harmonic coefficients
$C(l)$ and the autocorrelation function $N(\theta)$. The
estimator for $C(l)$ is defined as
\begin{equation}
  C(l)=\frac{1}{2l+1}\frac{1}{{\cal N}^2}
  \sum_{m=-l}^l\left(\sum_{i=1}^N\frac{1}{\omega_i}Y_{lm}(u^i)
  \right)^2\,,\label{cl}
\end{equation}
where $\omega_i$ is the total experimental exposure
at arrival direction $u^i$, ${\cal N}=\sum_{i=1}^{N}1/\omega_i$
is the sum of the weights $1/\omega_i$, and
$Y_{lm}(u^i)$ is the real-valued spherical harmonics function
taken at direction $u^i$. The estimator for $N(\theta)$ is defined as
\begin{equation}
N(\theta)=\frac{C}{S(\theta)}\sum_{j \neq i}
\left\{\begin{array}{ll}
1 & \mbox {if $\theta_{ij}$ is in same bin as $\theta$}\\
0 & \mbox{otherwise}
\end{array}\right\}\,,
\label{auto}
\end{equation}
and $S(\theta)$ is the solid angle size of the corresponding bin.
In Eq.~(\ref{auto}) the normalization factor $C=\Omega_e/(N(N-1))$,
with $\Omega_e$ denoting the solid angle of the sky region where the
experiment has non-vanishing exposure, is chosen such that an
isotropic distribution corresponds to $N(\theta)=1$.

The different mock data sets in the various realizations
yield the statistical distributions of $C(l)$ and $N(\theta)$.
One defines the average over all mock data sets and realizations
as well as two errors. The smaller error (shown to the left of
the average in the figures below) is the statistical
error, i.e. the fluctuations due to the finite number $N$ of observed
events, averaged over all realizations. The larger error
(shown to the right of the average in the figures below)
is the ``total error'', i.e. the statistical error plus the
cosmic variance. 
Thus, the latter includes the fluctuations due to finite
number of events and the variation between different realizations
of observer and source positions. 

Given a set of observed and simulated events, after extracting 
some useful statistical quantities $S_i$, 
namely $C_l$ and $N(\theta)$ defined above, we define
\begin{equation}
  \chi_n\equiv\sum_i
  \left(\frac{S_{i,{\rm data}}-\overline{S}_{i,{\rm simu}}}
             {\Delta S_{i,{\rm simu}}}\right)^n\,,\label{chi_n}
\end{equation}
where $S_{i,{\rm data}}$ refers to $S_i$ obtained from the real data, and
$\overline{S}_{i,{\rm simu}}$ and $\Delta S_{i,{\rm simu}}$ are the
average and standard deviations of the simulated data sets. 
This measure of deviation from the average prediction can be used to
obtain an overall likelihood for 
the consistency of a given theoretical model with an observed data 
set by counting the fraction of simulated data sets with
$\chi_n$ larger than the one for the real data.

\section{Results}

In the following we compare the results obtained for the 
simulated UHECR propagation experiments described above with
the observational results. In accord with what was outlined in
the previous section, the comparison is based
on the statistical properties of the simulated and observed
events, expressed in terms of the angular power spectrum and 
the autocorrelation function of the UHECR arrival
distributions. A summary of the simulations run is contained
in Tab.~\ref{tab}. There, for comparison, simulations 2 and 5
were performed for an observer situated in a small void with
weak ambient magnetic fields.

\begin{table}[h]
\caption[...]{List of UHECR propagation simulations. The columns contain
the simulation number, the number of sources in the simulation box
of $(50\,{\rm Mpc}\,h_{67}^-1)^3$, the magnetic field
strength at the observer location, the best fit power law index
in the injection spectrum $E^{-\alpha}$, and the overall likelihoods
of fits to the AGASA data above $4\times10^{19}\,$eV for the
multi-poles Eq.~(\ref{cl}) with $l\leq10$ and the auto-correlation
Eq.~(\ref{auto}) for $\theta\leq10^\circ$, respectively. The
likelihoods are computed for $n=4$ in Eq.~(\ref{chi_n}) which
leads to reasonable discriminative power.}\label{tab}
\begin{ruledtabular}
\begin{tabular}{cccccc}
\#&$N_s$&$B_{\rm obs}/$G&$\alpha$&${\cal L}_{l\leq10}$&
${\cal L}_{\theta\leq10^\circ}$\\
\hline \\
1 &100& $1.3\times10^{-7}$& 2.4&0.13 & 0.63 \\
2 &100& $8.2\times10^{-12}$& 2.7&0.098& 0.15\\
3 &10& $1.3\times10^{-7}$& 2.4&0.12 & 0.69 \\
4 &10& $2.7\times10^{-7}$& 2.4&0.071& 0.15\\
5 &10& $8.2\times10^{-12}$& 2.7& 0.011& 0.037\\
6 &1 & $1.3\times10^{-7}$& 2.8&0.074 & 0.62 \\
\end{tabular}
\end{ruledtabular}
\end{table}

\begin{figure}[ht]
\includegraphics[width=0.5\textwidth,clip=true]{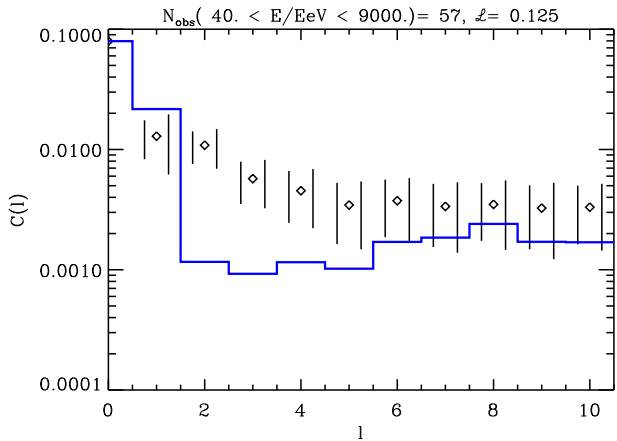}
\caption[...]{The angular power spectrum $C(l)$ as a function of
multi-pole $l$, obtained for the AGASA exposure function, see text,
for $N=57$ events observed above 40 EeV, sampled from 12 simulated
configurations of scenario 1 in Tab.~\ref{tab}. The diamonds indicate
the realization averages, and the left and right error
bars represent the statistical and total (including cosmic variance
due to different realizations)
error, respectively, see text for explanations. The histogram
represents the AGASA data. The overall likelihood significance is
$\simeq0.13$ for $n=4$ and $l\leq10$ in Eq.~(\ref{chi_n}).}
\label{fig2}
\end{figure}

\begin{figure}[ht]
\includegraphics[width=0.5\textwidth,clip=true]{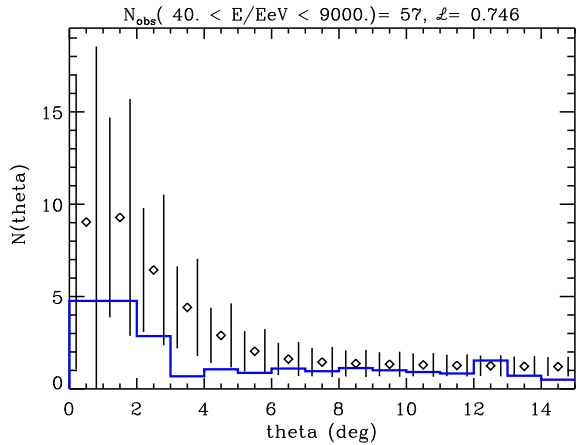}
\caption[...]{As Fig.~\ref{fig2}, but for the angular correlation
function $N(\theta)$ as a function of angular distance $\theta$,
using a bin size of $\Delta\theta=1^\circ$. Note that an isotropic
distribution would correspond to $N(\theta)=1$. The overall likelihood
significance is $\simeq0.63$ for $n=4$ and $\theta\leq10^\circ$ in
Eq.~(\ref{chi_n}). It is not significantly different for somewhat
larger bin sizes $\Delta\theta\simeq2^\circ$.}
\label{fig3}
\end{figure}

\begin{figure}[ht]
\includegraphics[width=0.5\textwidth,clip=true]{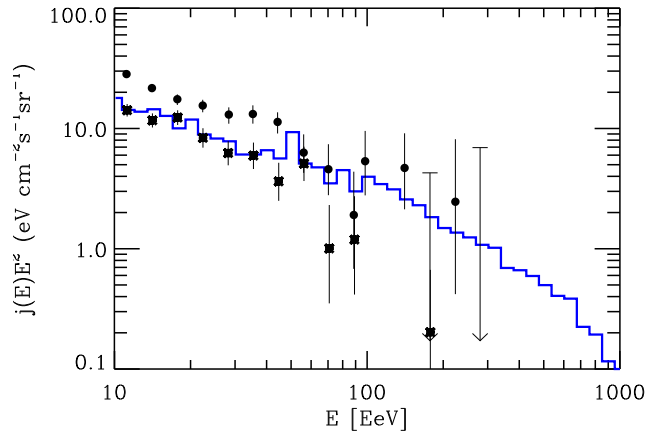}
\caption[...]{Predicted spectrum observable by AGASA for the scenario
1 in Tab.~\ref{tab}, for which multipoles and auto-correlations were
shown in Figs.~\ref{fig2} and~\ref{fig3}, averaged over 12 realizations,
as compared to the AGASA (dots) and HiRes-I (stars) data.}
\label{fig4}
\end{figure}

We find that as long as the observer is surrounded by magnetic
fields of about $0.1\mu$G, $N_s\gtrsim10$ nearby sources, i.e. sources
within the simulation box, are necessary
to reproduce multi-poles and autocorrelations marginally consistent
with present data, limited, we emphasize, to the Northern hemisphere only.
However, consistency of large scale multi-poles is somewhat worse
than for the spatially
more extended EGMF assumed in previous work~\cite{is}.
In Figs.~\ref{fig2} and~\ref{fig3} we show as an example the results
for the case of $N_s=100$ nearby sources, scenario 1 in Tab.~\ref{tab},
corresponding to a source density of
$8\times10^{-4}h_{67}^3\,\,{\rm Mpc}^{-3}$. The overall likelihood for $n=4$
in Eq.~(\ref{chi_n}) is $\simeq0.13$ and $\simeq0.63$ for the
multi-poles and autocorrelations
shown, respectively. Also Fig.~\ref{fig4} shows that, for 
UHECR sources characterized by a proton injection spectrum roughly 
as $\propto E^{-2.4}$ and extending up to $\simeq10^{21}\,$eV,
the observed spectrum at sub-GZK energies is well reproduced.
In addition, above GZK energies the spectral slope is predicted
to be somewhere between the AGASA and HiRes observations, see
Fig.~\ref{fig4}. Normalizing to the observed flux results
in a UHECR power of $5\times10^{41}{\rm erg}\,{\rm s}^{-1}$ per source
to be continuously emitted {\it above} $10^{19}\,$eV.

The situation for $N_s=10$ nearby sources does not lead to significantly
different likelihoods, see scenarios 3 and 4 in Tab.~\ref{tab}.
However, the case of just one source is clearly disfavored in terms
of the multi-poles, see scenario 6 in Tab.~\ref{tab}. This confirms
similar findings in earlier work~\cite{ils}.

\begin{figure}[ht]
\includegraphics[width=0.5\textwidth,clip=true]{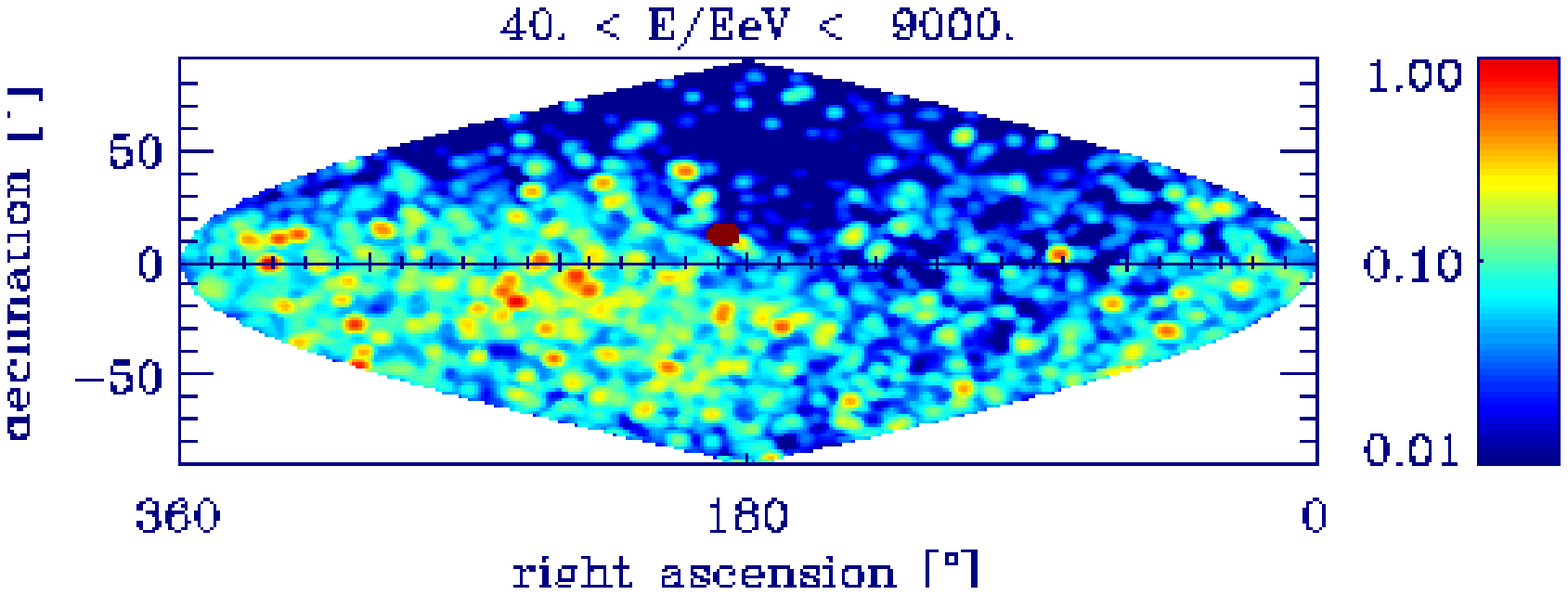}
\includegraphics[width=0.5\textwidth,clip=true]{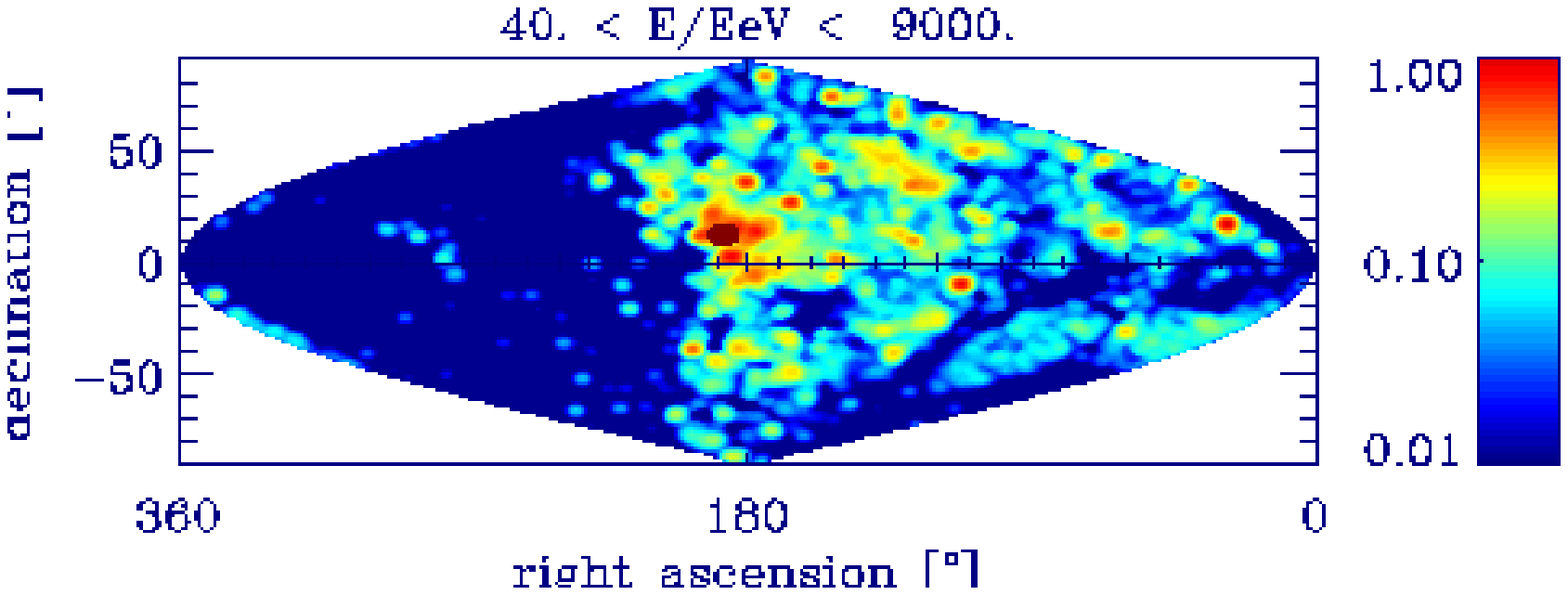}
\caption[...]{Illustration of the influence of magnetic fields
surrounding the observer on UHECR arrival direction distributions
above 40 EeV in terrestrial coordinates. The upper panel is for
scenario 1 (observer surrounded by relatively strong magnetic
fields), and the lower panel for scenario 2 (observer surrounded by
negligible magnetic fields) from Tab.~\ref{tab}, averaged over all 12
and 10 realizations of 5000 trajectories each, respectively, thus
corresponding to an effective number of sources of $\sim1000$. The
color scale represents the integral
flux per solid angle. The pixel size is $1^{\circ}$ and the image
has been convolved to an angular resolution of $2.4^{\circ}$
corresponding to the approximate AGASA angular
resolution. The filled sphere represents the position of the Virgo-like
cluster.}
\label{fig5}
\end{figure}

If the observer is in a region of EGMF strength much smaller
than $\simeq0.1\mu$G, as in scenario 2 of Tab.~\ref{tab}, for
$N_s\gtrsim100$ nearby sources the predicted
UHECR sky distribution reflects the highly structured large scale
galaxy distribution, smeared out only by the fields surrounding the
sources. This becomes obvious from Fig.~\ref{fig5} which shows
that UHECR arrival directions are much less isotropic in this case
than if the observer is immersed in fields $B\simeq0.1\mu$G.

\begin{figure}[ht]
\includegraphics[width=0.5\textwidth,clip=true]{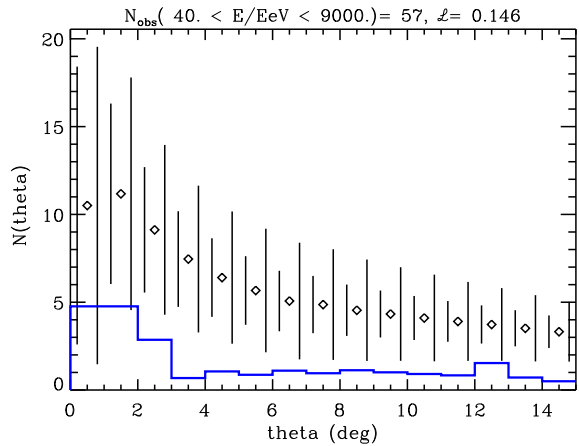}
\caption[...]{Similar to Fig.~\ref{fig3}, but for an observer
in a much lower field region, $\simeq8.2\times10^{-12}\,$G.
This corresponds to scenario 2 in Tab.~\ref{tab}.
The overall likelihood significance is $\simeq0.15$ for $n=4$
and $\theta\leq10^\circ$ in Eq.~(\ref{chi_n}).}
\label{fig6}
\end{figure}

Nevertheless, the overall likelihood significance for multi-poles
up to $l=10$ is $\simeq0.1$, and thus not significantly worse than
for the strong observer field case of Fig.~\ref{fig2}. 
Therefore, the number of events observed by AGASA above $40\,$EeV
is insufficient to distinguish this low observer field case from
the strong observer field case based on anisotropy alone.
However, as can be seen from Fig.~\ref{fig6},
the low observer field case results in auto-correlations 
at angles $\theta\gtrsim3^\circ$ much larger than observed by AGASA.
This is because strong magnetic fields at the observer position
cause enough UHECR diffusion that their large-scale auto-correlations
are significantly suppressed, as in Fig.~\ref{fig3}. However,
for fields considerably larger than $0.1\mu$G the auto-correlations
tend to become too strong again, see scenario 4 in Tab.~\ref{tab},
probably due to increased magnetic lensing.

\begin{figure}[ht]
\includegraphics[width=0.5\textwidth,clip=true]{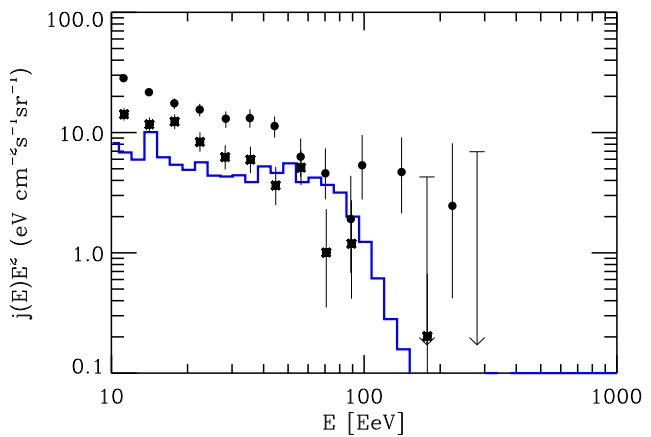}
\caption[...]{The unnormalized energy spectrum observed at Earth resulting
from a $E^{-2.4}$ isotropic proton flux injected at a sphere
of 40 Mpc radius around the observer for scenario 1 from Tab.~\ref{tab},
averaged over solid angle and $10^4$ computed trajectories. In
contrast to the spectrum of the local component shown in Fig.~\ref{fig4},
there is a clear tendency that this cosmological component can fit the
flux neither at the highest nor at the lowest energies.}
\label{F6}
\end{figure}

We also find that sources outside our Local Supercluster
do not contribute significantly to the observable flux if the observer
is immersed in magnetic fields above about $0.1\mu$G and if
the sources reside in magnetized clusters and super-clusters:
For particles above the GZK cutoff this is because sources outside
the Local Supercluster are beyond the GZK distance. 
On the other hand, sub-GZK particles are mainly confined in their local
magnetized environment 
and thus exhibit a much higher local over-density than
their sources. Further, the suppressed flux of low energy particles
leaving their environment is largely kept away from the observer
if he is surrounded by significant magnetic fields~\cite{is}. 
Both effects can be
understood qualitatively by matching the flux $j(E)$ in the unmagnetized
region with the diffusive flux $-D(E){\bf\nabla}n(E,{\bf r})$ in terms
of the diffusion coefficient $D(E)$ and the density $n(E,{\bf r})$ of
particles of energy $E$ which shows that the density gradient always
points to the source.
More quantitatively, the shape of the large-distance
component is demonstrated in Fig.~\ref{F6} which shows the observable
flux resulting from an $E^{-2.4}$ spectrum injected isotropically
at a sphere with a radius of 40 Mpc around the observer. Note that
despite the smaller energy losses the sub-GZK particles arriving from outside
the Local Supercluster are likely to have a spectrum even more strongly
suppressed than in Fig.~\ref{F6} at low energies due to their
containment in the source region. A significant contribution from
sources at cosmological distances to sub-GZK energies thus requires
that neither these sources nor the observer are immersed in too
strong magnetic fields and/or an injection spectrum considerably
steeper than $E^{-2.4}$ to compensate for the systematic suppression
of flux of lower energy particles.

\begin{figure}[ht]
\includegraphics[width=0.5\textwidth,clip=true]{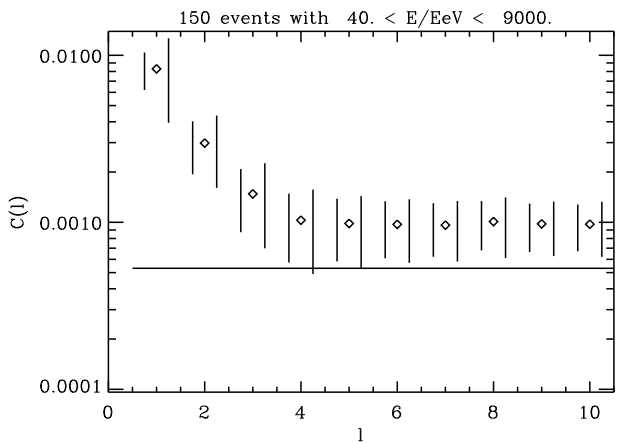}
\caption[...]{Same as Fig.~\ref{fig2}, but for comparison of
the model predictions with an isotropic distribution [horizontal line,
$C_l\simeq(4\pi N)^{-1}$, see Eq.~(\ref{cl})]
for the full-sky detector \`a la Auger discussed in the text,
for $N=150$ events observed above 40 EeV.}
\label{fig8}
\end{figure}

The confidence levels that can be obtained with this method for
specific models of our local magnetic and UHECR source neighborhood
will greatly increase with the increase of data from future
experiments. Full sky coverage alone will play an important role
in this context as many scenarios predict large dipoles for the
UHECR distribution. This is the case for basically all scenarios
considered here, as demonstrated in Fig.~\ref{fig8}. Whereas current
northern hemisphere data are consistent with scenarios with
$N_s\gtrsim10$ nearby sources at the $\simeq1.5\,$sigma level
if the observer is surrounded by relatively strong fields
$B\sim0.1\,\mu$G, a comparable or larger exposure in the southern
hemisphere would be sufficient in these cases to find a dipole at
several sigma confidence level, as demonstrated in Fig.~\ref{fig8}.

\begin{figure}[ht]
\includegraphics[width=0.5\textwidth,clip=true]{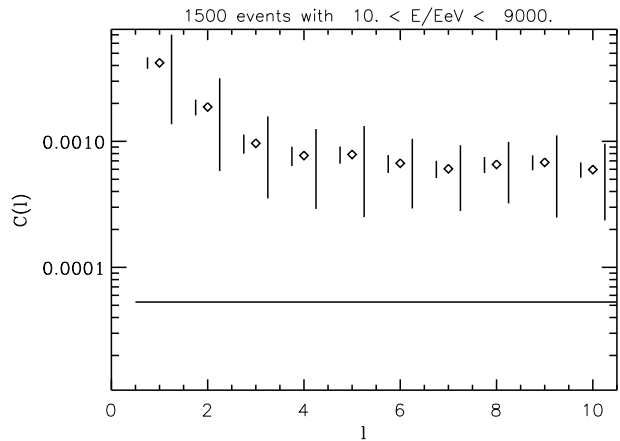}
\caption[...]{Same as Fig.~\ref{fig8}, but for
$N=1500$ events observed above 10 EeV.}
\label{fig9}
\end{figure}

Finally, the distributions of events down to $10^{19}\,$eV also
contain important information. Fig.~\ref{fig9} shows the multi-poles
predicted by our standard scenario 1 in Tab.~\ref{tab} that
full-sky experiment would observe for 1500 events detected above
$10^{19}\,$eV. This corresponds to twice the number of currently
observed AGASA events and thus approximately reflects the current
exposure. A corresponding figure for the AGASA detector
alone would look similar. It is obvious that there is significant anisotropy
even at $l\simeq10$, inconsistent with current AGASA observations.
On the other hand, cosmic variance becomes more
important at these lower energies, and a possible significant
contribution from large-distance sources cannot
be excluded if their magnetization is not too high, as discussed
above. It is easy to see from Eq.~(\ref{cl}) that if a fraction
$f_a$ of $N$ events observed stems from an anisotropic, local
contribution, whereas the fraction $1-f_a$ is cosmological and
completely isotropic, then
\begin{equation}
  C_l\simeq C_{l,i}\left((1-f_a)^2+\frac{C_{l,a}}{C_{l,i}}f_a^2\right)
  \,,\label{part}
\end{equation}
where $C_{l,i}=(4\pi N)^{-1}$ and $C_{l,a}$ are the expectation values
of $C_l$ for the isotropic and the anisotropic distribution,
respectively.  Therefore, at $\simeq10^{19}\,$eV an isotropic
cosmological flux about a factor 3 higher than the anisotropic flux
originating within $\simeq50\,$Mpc would be needed to explain the
isotropy observed by AGASA. For charged primaries this implies steep
injection spectra and/or weak magnetic fields around observer and
sources, as explained above. Without going into a more detailed
analysis we remark that this will also require to decrease the flux
contribution from nearby sources shown in Fig.~\ref{fig4} at the low
energy end. As a consequence, the best fit injection spectrum for the
local component will be slightly harder than the power law indices
shown in Tab.~\ref{tab}.  This is consistent with what is expected
from shock acceleration theory~\cite{rel_shock}.

\section{Conclusions}
In the present work we performed UHECR propagation simulations based
on the distributions of magnetic field and baryon density obtained
from a simulation of large scale structure formation. The magnetic
field was simulated as a passive quantity and normalized at simulation
end in agreement with published measurements of Faraday rotation 
measures for groups and clusters of galaxies~\cite{bo_review}.
We considered finite numbers of discrete UHECR sources with equal
total power and injection spectrum. Their positions
were randomly selected with probability proportional to the baryon
density. The observer was chosen within small groups of galaxies 
characterized by gas temperatures around a fraction of a keV,
typical for our local environment. One chosen observer was found
in a relatively high field region with $B\simeq0.1\mu$G. For
comparison, we also chose an observer situated in a small void,
where the surrounding field is $B\simeq10^{-11}\,$G. 
We found that good fits to the AGASA data above $4\times10^{19}\,$eV
in the Northern hemisphere are only obtained for
$N_s\gtrsim10$ sources and for observers surrounded by $\simeq0.1\mu$G
fields. Otherwise the predicted arrival direction distribution is
either too anisotropic or produces too large auto-correlations at angles
larger than a few degrees. The best fit case occurs for $N_s\simeq100$,
significantly higher than in previous work~\cite{is} due to the
more localized and more strongly structured magnetic fields considered
here.

For the required local source number density and continuous power per
source we find $n_{\rm source} \gtrsim10^{-4}-10^{-3}\,h_{67}^3{\rm
Mpc}^{-3}$, and $Q_{\rm source} \lesssim 5\times10^{41}{\rm erg}\,{\rm
s}^{-1}$ respectively, the latter within about one order of magnitude
uncertainty to both sides.  This corresponds to an average UHECR
emissivity of $q_{\rm UHECR} =n_{\rm source}\, Q_{\rm source} \sim
10^{38} {\rm erg\,Mpc^{-3}\,s^{-1}}$ also with an uncertainty of
roughly one order of magnitude, not larger,
since it is fixed by the observed UHECR flux.

Possible sources marginally consistent with these energy requirements
are radio galaxies.  Their present energy release of $q_{\rm rg} \sim
5\times 10^{39}\,{\rm erg\,s^{-1}\,Mpc^{-3}} \,(f_{\rm power}/10)$
\cite{ebkw97} is roughly what is required in order to produce a
sufficient flux of UHECR, assuming that the injection power law is
flat ($\propto E^{-2}$) \cite{rb93,rsb93}.  The parameter $f_{\rm
power}$ describes the ratio of the total power of the radio galaxy to
the equipartition estimate based on its radio luminosity, and it
enters the used radio luminosity-jet-power relation of \cite{ebkw97}.
We expect $f_{\rm power} \sim 10$ within an order of magnitude. In the
estimate of the radio galaxy power the observed radio luminosity
function~\cite{dp90} was integrated only for sources with a 2.7 GHz
luminosity of more than $L_{\rm min} = 2\times 10^{22}\,{\rm
Watt\,Hz^{-1}}$, since they correspond to a luminosity of
$10^{43}\,{\rm erg\,s^{-1}}\,(f_{\rm power}/10)$. This would provide the
required UHECR luminosity per source of $Q_{\rm source} \sim
5\times10^{41}{\rm erg}\,{\rm s}^{-1} (f_{\rm power}\,f_{\rm UHECR}
/0.5)$, using the optimistic assumption of \cite{rb93,rsb93} that
$f_{\rm UHECR} = 3-10\%$ of the radio galaxy power is converted into
UHECRs. The implied number density of these radio galaxies is $n_{\rm
source} \sim 10^{-4}\,{\rm Mpc^{-3}}$ and, therefore, is only marginally
consistent with the required $n_{\rm source} \sim
10^{-4}-10^{-3}\,h_{67}^3{\rm Mpc}^{-3}$. Since the number density
increases strongly with decreasing $L_{\rm min}$ this requirement can
possibly be fulfilled by allowing for a larger number of less powerful
UHECR sources. This implies that basically every radio galaxy has to
be an efficient UHECR source, not only the most powerful ones. Since
many of the weaker radio galaxies do not exhibit a hot spot, which is
assumed to be the UHECR acceleration site in the scenario of
\cite{rb93,rsb93}, their efficiency in producing UHECR might be largely
reduced. This is a potential serious problem for this scenario, since
lowering $f_{\rm UHECR}$ by several orders of magnitude can not be
fully compensated by assuming a higher radio galaxy jet-power, because
$f_{\rm power} > 100$ does not seem to be consistent with observations
of radio galaxies \cite{1992A&A...265....9F}.

To conclude, radio galaxies can be the sources of UHECRs if even weak
radio galaxies are efficient particle accelerators to the highest
energies, otherwise they have serious problems to reproduce the 
smooth UHECR arrival direction distribution.

We also found that consistency with the isotropy observed by AGASA
down to $10^{19}\,$eV requires the existence of an isotropic
component with a flux about a factor 3 larger than the local
component. This isotropic component would presumably be of
cosmological origin and thus would not contribute significantly
above $4\times10^{19}\,$eV due to the GZK effect, consistent with
the fact that at these energies we find local scenarios consistent
with all data. The resulting best fit injection spectrum for the
local component is $E^{-(2.2-2.4)}$. In contrast, for the charged primaries
of the cosmological component to dominate around $10^{19}\,$eV
steep injection spectra and/or weak magnetic fields around
observer and sources would be required. These two conflicting requirements 
provide a strong argument against the hypothesis
of local astrophysical sources of UHECRs above $\simeq10^{19}\,$eV
in a strongly magnetized and structured intergalactic medium.

Finally, we have also demonstrated that already a modest increase in data
together with full-sky coverage will allow to put considerably
stronger constraints on UHECR source and magnetic field scenarios
than presently possible. In particular, our local scenarios predict
the emergence of significant dipoles and quadrupoles above
$4\times10^{19}\,$eV.

Modeling our cosmic neighborhood and simulating UHECR propagation in
this environment will therefore become more and more important in the
coming years. This will also have to include the effects of the
Galactic magnetic field and an extension to a possible heavy component
of nuclei. For first steps in this direction see,
e.g. Refs.~\cite{mte,ames}, and Ref.~\cite{bils}, respectively.

\section*{Acknowledgments}
We would like to thank Martin Lemoine and Claudia Isola for
earlier collaborations on the codes partly used in this work
and to F.~W.~Stecker for useful comments to the manuscript.
The work by FM was partially supported by the Research and 
Training Network ``The Physics of the Intergalactic Medium''
set up by the European Community
under the contract HPRN-CT2000-00126 RG29185.
The computational work was carried out at the
Rechenzentrum in Garching operated by the 
Institut f\"ur Plasma Physics
and the Max-Planck Gesellschaft.

\end{document}